\documentclass[screen,nonacm,sigconf]{acmart}
\AtBeginDocument{%
  }

\copyrightyear{2023} 

\usepackage{listings}
\usepackage{color}

\definecolor{gray}{rgb}{0.4,0.4,0.4}
\definecolor{darkblue}{rgb}{0.0,0.0,0.6}
\definecolor{cyan}{rgb}{0.0,0.6,0.6}

\lstdefinelanguage{XML}
{
  morestring=[b]",
  morestring=[s]{>}{<},
  morecomment=[s]{<?}{?>},
  stringstyle=\color{black},
  identifierstyle=\color{darkblue},
  keywordstyle=\color{cyan},
  morekeywords={xmlns,version,type},
  showstringspaces=false
}

\begin{document}

\title{Scaling Computational Fluid Dynamics: In Situ Visualization of NekRS using SENSEI}

\author{Victor A. Mateevitsi}
\affiliation{%
  \institution{Argonne National Laboratory}
  \country{United States of America}
}
\email{vmateevitsi@anl.gov}

\author{Mathis Bode}
\affiliation{%
  \institution{Forschungszentrum J\"ulich \\ J\"ulich Supercomputing Centre}
  \country{Germany}
}
\email{m.bode@fz-juelich.de}

\author{Nicola Ferrier}
\affiliation{%
  \institution{Argonne National Laboratory}
  \country{United States of America}
}
\email{nferrier@anl.gov}

\author{Paul Fischer}
\affiliation{%
  \institution{Argonne National Laboratory}
  \institution{University of Illinois Urbana-Champaign}
  \country{United States of America}
}
\email{fischerp@illinois.edu}

\author{Jens Henrik G\"obbert}
\affiliation{%
  \institution{Forschungszentrum J\"ulich \\ J\"ulich Supercomputing Centre}
  \country{Germany}
}
\email{j.goebbert@fz-juelich.de}

\author{Joseph A. Insley}
\affiliation{%
  \institution{Argonne National Laboratory}
  \institution{Northern Illinois University}
  \country{United States of America}
}
\email{insley@anl.gov}

\author{Yu-Hsiang Lan}
\affiliation{%
  \institution{Argonne National Laboratory}
  \institution{University of Illinois Urbana-Champaign}
  \country{United States of America}
}
\email{ylan@anl.gov}

\author{Misun Min}
\affiliation{%
  \institution{Argonne National Laboratory}
  \country{United States of America}
}
\email{mmin@anl.gov}

\author{Michael E. Papka}
\affiliation{%
  \institution{Argonne National Laboratory}
  \institution{University of Illinois at Chicago}
  \country{United States of America}
}
\email{papka@anl.gov}

\author{Saumil Patel}
\affiliation{%
  \institution{Argonne National Laboratory}
  \country{United States of America}
}
\email{spatel@anl.gov}

\author{Silvio Rizzi}
\affiliation{%
  \institution{Argonne National Laboratory}
  \country{United States of America}
}
\email{srizzi@anl.gov}

\author{Jonathan Windgassen}
\affiliation{%
  \institution{Forschungszentrum J\"ulich \\ J\"ulich Supercomputing Centre}
  \country{Germany}
}
\email{j.windgassen@fz-juelich.de}

\renewcommand{\shortauthors}{Mateevitsi et al.}


\begin{abstract}In the realm of Computational Fluid Dynamics (CFD), the demand for memory and computation resources is extreme, necessitating the use of leadership-scale computing platforms for practical domain sizes. This intensive requirement renders traditional checkpointing methods ineffective due to the significant slowdown in simulations while saving state data to disk. As we progress towards exascale and GPU-driven High-Performance Computing (HPC) and confront larger problem sizes, the choice becomes increasingly stark: to compromise data fidelity or to reduce resolution. To navigate this challenge, this study advocates for the use of \textit{in situ} analysis and visualization techniques. These allow more frequent data "snapshots" to be taken directly from memory, thus avoiding the need for disruptive checkpointing. We detail our approach of instrumenting NekRS, a GPU-focused thermal-fluid simulation code employing the spectral element method (SEM), and describe varied \textit{in situ} and in transit strategies for data rendering. Additionally, we provide concrete scientific use-cases and report on runs performed on Polaris, Argonne Leadership Computing Facility's (ALCF) 44 Petaflop supercomputer and Jülich Wizard for European Leadership Science (JUWELS) Booster, Jülich Supercomputing Centre's (JSC) 71 Petaflop High Performance Computing (HPC) system, offering practical insight into the implications of our methodology.
\end{abstract}



\maketitle

\section{Introduction}
Rooted in the Spectral Element Method (SEM) \cite{fischer_spectral_1988}, NekRS \cite{fischer_nekrs_2022} is a GPU-accelerated thermal-fluid simulation code. Rapidly becoming a staple in modeling and simulating turbulent flows, it finds uses across a wide spectrum, from simulating internal combustion engines to large-scale atmospheric and oceanic flows, as well as reactor thermal hydraulics. The challenges posed by these simulations are significant in terms of scale, resolution, and computational demand. NekRS tackles these hurdles using libParanumal \cite{karakus_gpu_2019, swirydowicz_acceleration_2019}, a high-order finite element solver, and the OCCA library \cite{medina_occa_2014, medina_okl_2015}, a hardware-agnostic solution, to enable optimized performance and scalability.

With the advent of exascale supercomputers like Argonne National Laboratory's Aurora \cite{stevens_aurora_2019}, the disparity between rapid on-chip processing and slower disk storage is set to widen. Data-saving to disk could notably hamper simulations, necessitating pauses for I/O operations to complete. This situation leaves scientists with a tough choice: reduce checkpoint frequency or simplify the domain by lowering the resolution, both potentially resulting in overlooked discoveries.

\textit{In situ} processing \cite{ma_-situ_2007}, which facilitates data processing while it remains in memory, presents a compelling solution to this challenge. The SENSEI project \cite{ayachit_sensei_2016} embodies this approach, aiming to equip the simulation code with the flexibility to interchange \textit{in situ} algorithms without recompilation. In this paper, we demonstrate the instrumentation of NekRS with SENSEI and present our scalability experiments conducted using real scientific use-cases.

\section{Background}
The advancement in computational capabilities has significantly outpaced that of I/O speeds, leading to a heightened interest in \textit{in situ} processing. This technique revolves around the immediate analysis and visualization of data, directly from memory, sidestepping the delays of I/O operations. As a result, simulations become more efficient with reduced cycles spent on intensive I/O. Early adoptions of this method can be traced back to Zajac et al. \cite{zajac_computer-made_1964}, who showcased a simulation of Earth and satellite orbits plotted straight onto microfilm. However, the rise of heterogeneous systems and the sheer variety of analysis algorithms and graphics Application Programming Interfaces (APIs) like OpenGL, DirectX, Vulkan, and ANARI have compounded the intricacy of embedding universal cross-platform \textit{in situ} capabilities in scientific codes. Dedicated \textit{in situ} libraries are now essential, offering specialized tools for such diverse platforms, simplifying tasks such as image rendering, and easing code instrumentation.

\subsection{In situ libraries}
Presently, several key libraries, notably Ascent \cite{larsen_ascent_2022}, Catalyst \cite{ayachit_paraview_2015}, LibSim \cite{kuhlen_parallel_2011, childs_situ_2012}, and SENSEI \cite{ayachit_sensei_2016} are the most active ones. Ascent, part of the ALPINE project and backed by the U.S. Department of Energy's (DOE) Exascale Computing Project (ECP) \cite{messina_exascale_2017}, is a lightweight \textit{in situ} visualization and analysis library built for multi-physics HPC simulations. Its unique selling point is its minimal reliance on external dependencies, with VTK-m \cite{moreland_vtk-m_2016} employed for rendering. Catalyst, maintained by Kitware, relies on VTK, facilitating detailed visualization workflows via VTK's comprehensive visualization tools. In a similar vein, LibSim \cite{kuhlen_parallel_2011, childs_situ_2012} operates alongside VisIt \cite{childs_visit_2012}.

\subsection{Computational Fluid Dynamics}
Computational Fluid Dynamics (CFD) employs numerical methods to study fluid flow problems \cite{anderson_computational_1995}. Its applications span from designing combustors and simulating nuclear reactors to aerodynamic modeling of aircrafts and space shuttles \cite{khalil_cfd_2021}. Nek5000, pioneered in the late 80s \cite{fischer_spectral_1988, fischer_spectral_1989}, stands out as a gold standard for spectral element simulations, scaling efficiently across varying computational platforms. Instrumentation studies by Atzori et al. \cite{atzori_situ_2022} and Bernardoni et al. \cite{bernardoni_situ_2018} shed light on its adaptability with \textit{in situ} libraries like Catalyst and SENSEI. Despite Nek5000's enduring prominence over 35 years, the advent of GPU-accelerated HPCs necessitated a major code overhaul, giving rise to its successor, NekRS \cite{fischer_nekrs_2022}.

\section{Methodology}
SENSEI is structured around two primary components: the AnalysisAdaptor and DataAdaptor.

\subsection{AnalysisAdaptor}
The \lstinline{AnalysisAdaptor} serves as the interface for \textit{in situ} analysis codes and algorithms, connecting with tools like Catalyst, VTK-m, or OSPray \cite{wald_ospray_2017}. Developing a new \lstinline{AnalysisAdaptor} necessitates extending the existing interface and implementing the desired analysis functionality. These adaptors offer flexibility; they can be swapped dynamically at runtime via an \lstinline{.xml} configuration file. For instance, to enable \textit{in situ} image rendering through Catalyst, one can activate the Catalyst \lstinline{AnalysisAdaptor} without needing to recompile the entire codebase, as illustrated in Listing \ref{lst:AnalysisAdaptor}.

\begin{lstlisting}[language=XML, frame=lines, caption={SENSEI AnalysisAdaptor Configuration}, captionpos=b, label={lst:AnalysisAdaptor}]  
<sensei>
  <analysis type="catalyst" pipeline="pythonscript" filename="analysis.py" frequency="100" />
</sensei>
\end{lstlisting}

\subsection{DataAdaptor}
Data is channeled to the \lstinline{AnalysisAdaptor} through the use of the \lstinline{DataAdaptor}. This component is a C++ interface provided by SENSEI, which simulation codes must extend and implement. The role of this child adaptor is to relay data (aligned with the VTK data model) to SENSEI. Moreover, an intermediary "bridge" code is responsible for embedding SENSEI into the simulation, initializing the library, updating the data, managing adaptors, and periodically invoking analysis routines.

\begin{lstlisting}[language=C, frame=lines, caption={Pseudocode of the DataAdaptor}, captionpos=b, label={lst:DataAdaptor}]
class DataAdaptor : sensei::DataAdaptor 
{
  void Initialize(nek_data) { }
  int GetNumberOfMeshes() { };
  int GetMeshMetadata() { };
  int GetMesh() { };
  int AddArray() { };
};
\end{lstlisting}

Given the identical data models employed by both Nek5000 and NekRS, we crafted a unique \lstinline{nek_sensei::DataAdaptor} class (Listing \ref{lst:DataAdaptor}) and corresponding bridge code (Listing \ref{lst:Bridge}). To promote reusability and ease maintenance, we housed this code in a separate repository and integrated it into both Nek5000 and NekRS using a GitHub submodule. 

NekRS employs \lstinline{OCCA} \cite{medina_occa_2014}, a flexible, vendor-agnostic framework for parallel programming across heterogeneous platforms. When compiled with a GPU backend, such as CUDA, NekRS operates on the GPU. This poses a challenge, as simulation data residing on GPU device memory must be transferred to the CPU before being relayed to SENSEI due to VTK data model's current lack of GPU device memory support.

\begin{lstlisting}[language=C, frame=lines, caption={Pseudocode of the bridge code}, captionpos=b, label={lst:Bridge}]
void initialize(MPI_Comm* comm, nek_data) {
  NekDataAdaptor *da = new NekDataAdaptor();
  da->Initialize(nek_data);

  ConfigurableAnalysis ca = new ConfigurableAnalysis();
  ca->Initialize("conf.xml");
}

void update(double* t, DataAdaptor **d) { }
\end{lstlisting}

For the evaluation, we compiled NekRS latest version (v23), and used SENSEI's development branch. To enable the Catalyst AnalysisAdaptor, we compiled against ParaView 5.11.1, which was compiled with OSPRay support.

\section{Results}

To evaluate the effectiveness of the instrumentation, we conducted experiments on two distinct HPCs: Polaris, a 44 Petaflops HPE Cray GPU-based HPC, and JUWELS Booster, a 71 Petaflops Atos GPU-based HPC. Polaris houses 560 nodes, each fitted with a single AMD EPYC "Milan" processor and four NVIDIA A100 GPUs. JUWELS Booster consists of 936 compute nodes, each equipped with 2 AMD EPYC "Rome" CPUs and 4 NVIDIA A100 GPUs. The network has a DragonFly+ topology with HDR-200 InfiniBand.

The primary goal of these experiments is to quantify the computational overhead introduced by our workflow. The first use case run on Polaris and is an example for an \textit{in situ} application of our workflow, while the second one run on JUWELS Booster and uses an in transit setup.

To maintain comparability, we show similar performance metrics for both cases. However, we do not perform exactly the same analyses, as the focus is on providing deep insight into the application of different workflow strategies, rather than a one-to-one comparison of the supercomputing environments. For the sake of reproducibility, we have made all source and analysis code, use cases, and data available \cite{mateevitsi_software_2023}.




\subsection{In situ Pebble-bed reactor case}

\begin{figure}[h]
\includegraphics[width=0.8\linewidth]{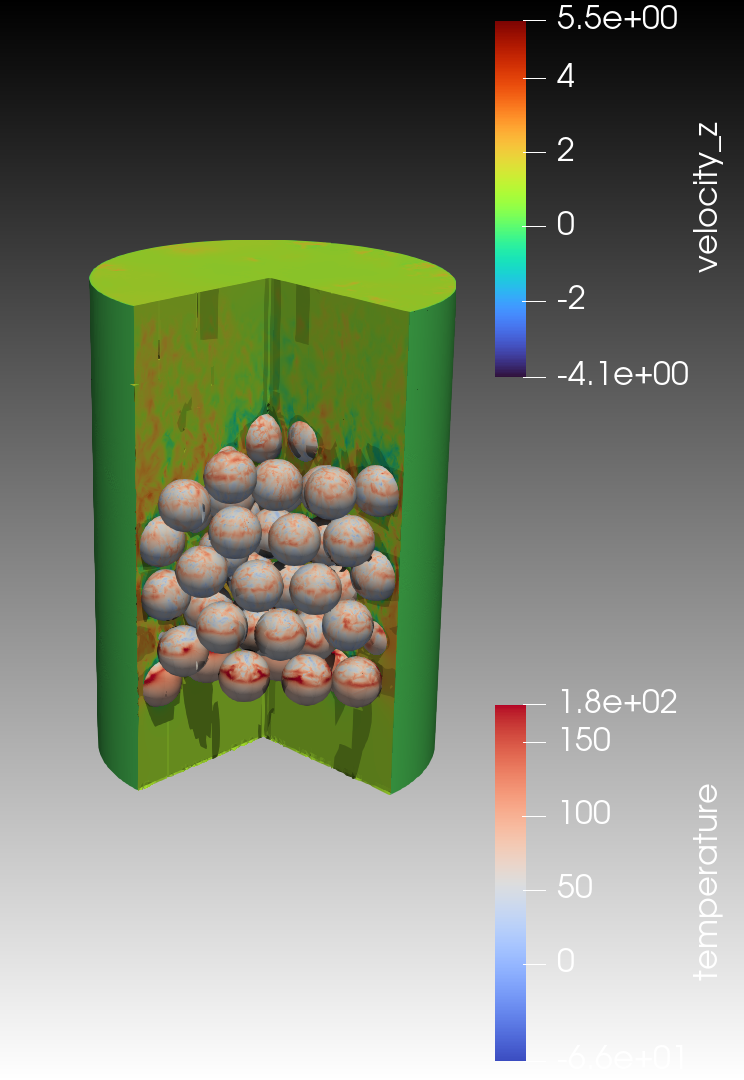}
\caption{Visualization of the pb146 use case simulation, illustrating flow dynamics within a pebble-bed nuclear reactor.}
\label{fig:pb146_visual}
\end{figure}

Our assessment hinged on two pivotal metrics: runtime and memory footprint. Herein, runtime refers to the total elapsed wall-clock time, while memory footprint corresponds to the aggregate memory high water mark across all MPI ranks. The configuration adopted for this test include:

\begin{itemize}
\item \textbf{Original:} Here, NekRS runs sans the SENSEI interface, serving as the baseline.
\item \textbf{Checkpointing:} This configuration witnesses NekRS run with built-in checkpointing initiated at every n frames. In this context, checkpointing pertains to the practice of periodically storing raw simulation data onto disk.
\item \textbf{Catalyst:} In this mode, NekRS, integrated with the SENSEI interface, leverages the Catalyst AnalysisAdaptor. Data is copied from the GPU to the CPU and subsequently passed to SENSEI, which employs the Catalyst Adaptor for rendering tasks.
\end{itemize}

The test bench used to test our instrumentation was the "pb146" use case simulation, an inherent example within the NekRS suite. This simulation models a computational fluid dynamics representation of a pebble-bed nuclear reactor core, housing 146 spherical pebbles (Figure \ref{fig:pb146_visual}), and runs on the GPUs. Such a simulation is of particular interest, given the growing interest in advanced carbon-neutral nuclear fission reactors \cite{min_optimization_2022}. For all configurations, we allowed the simulation to execute for 3,000 timesteps, instigating either checkpointing or \textit{in situ} processes at 100 timestep intervals.

\begin{figure}[h]
\includegraphics[width=0.7\linewidth]{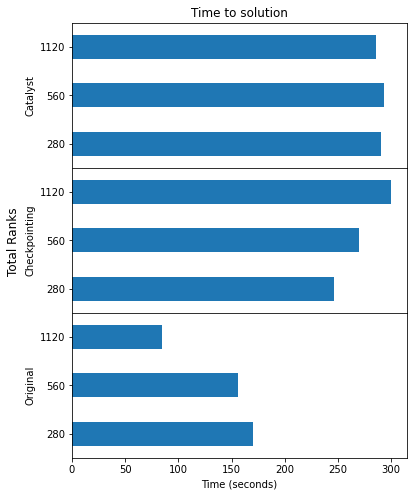}
\caption{Comparison of time-to-solution across 280, 560 and 1,120 rank runs for Catalyst, Checkpointing, and Original configurations.}
\label{fig:alcf_time_to_solution}
\end{figure}

\begin{figure}[h]
\includegraphics[width=0.8\linewidth]{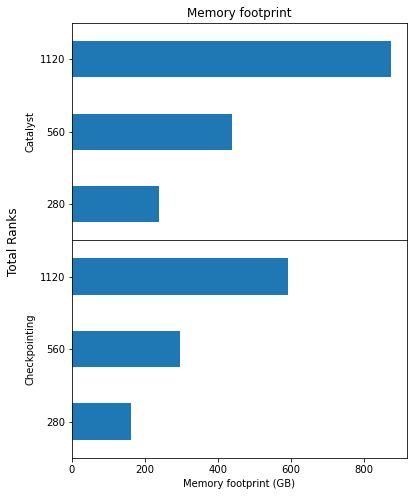}
\caption{Memory usage comparison between the 280, 560 and 1,120 rank runs for both Catalyst and Checkpointing configurations.}
\label{fig:alcf_memory_footprint}
\end{figure}

Trials were conducted on 70 nodes (12.5\% of Polaris, constituting 280 ranks), 140 nodes (25\% of Polaris, which is 560 ranks), and 280 nodes (50\% of Polaris, which is 1120 ranks) under both Catalyst and Checkpointing configurations. The elapsed time for the Original configuration was deduced by subtracting the Checkpointing time from the cumulative elapsed time. The outcome is depicted in Figure \ref{fig:alcf_time_to_solution}. As expected, the Original configuration showcased optimal time efficiency, unburdened by I/O or \textit{in situ} processing overheads. In juxtaposition, the Catalyst approach bore a slight overhead when pitted against Checkpointing. However, it's crucial to highlight that the storage demand for Catalyst was a mere 6.5MB, in stark contrast to the whopping 19GB necessitated by Checkpointing. This signifies that, while the computational overheads of \textit{in situ} almost mirror those of Checkpointing, they accomplish this at an impressive storage economy, nearly three orders of magnitude less. Delving deeper, Figure \ref{fig:alcf_memory_footprint} illustrates that Catalyst's CPU memory overhead is approximately 25\% greater. This escalation is rational, given the need to transition data from GPU to CPU and the inherent overhead accompanying Catalyst operations.

\subsection{In transit Mesoscale case}

\begin{figure}[h]
\includegraphics[width=\linewidth]{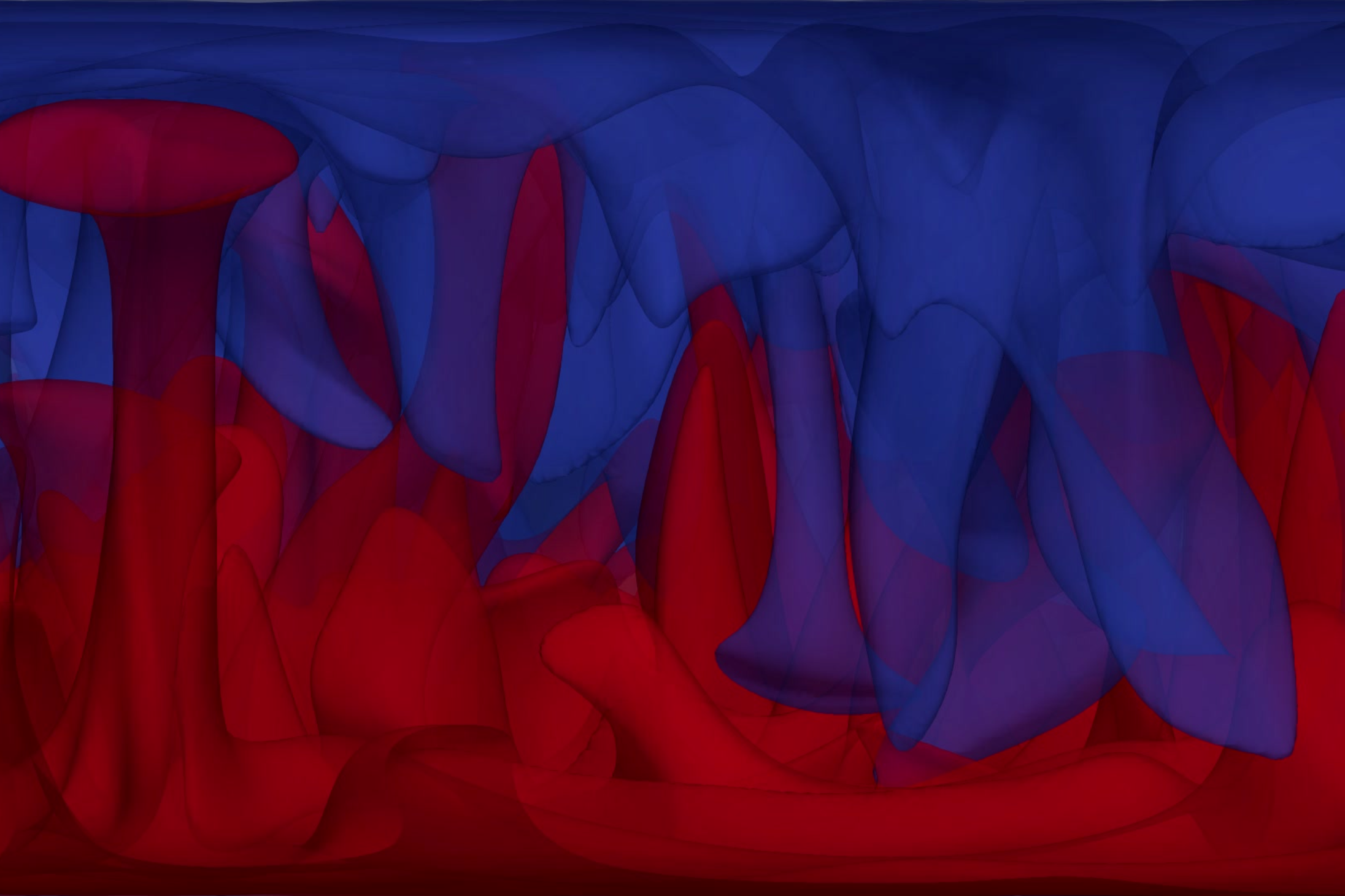}
\caption{Side view visualization of a RBC case.}
\label{fig:rbc}
\end{figure}

Rayleigh-Bénard convection (RBC) is one of the classical natural convection types in fluid thermodynamics and has been widely studied (Fig.~\ref{fig:rbc}). A basic setup leading to RBC is a fluid heated from below (i.e., perpendicular to the direction of gravity). Depending on the so-called Rayleigh number (Ra), the heat transfer in such a setup is dominated either by conduction (low Ra) or convection (high Ra), and a so-called Bénard cell is formed.

Numerically, high Ra setups are very interesting because these flows are highly turbulent, leading to a large separation of scales. Therefore, large meshes with high resolution are required and simulations are only possible with supercomputers, making it a perfect example for our workflow. A recent example of RBC computed on supercomputers used parameters relevant to conditions at the Sun's surface, resulting in so-called mesoscale convection \cite{pandey_convective_2022}.

For our scaling measurements, NekRS-SENSEI is complemented by ADIOS2 (v.2.9.1) for data transport, resulting in an in transit visualization workflow. A major advantage of the in-transit workflow is that the memory available for simulation nodes is independent of the number of visualization cores/nodes. Since available memory is often one of the limiting parameters for simulation size, this avoids unnecessary trade-offs between simulation size and visualization speed. The endpoint of our workflow is always a SENSEI data consumer, and the ratio of simulation nodes to endpoint nodes is 4:1 in all cases.

The Sustainable Staging Transport (SST) engine with its classic streaming data architecture is selected as the ADIOS2 engine.
It is configured to communicate via UCX for data transport and is set to use TCP sockets on Infiniband for control operations and BP as a data marshaling option.

This in-transit workflow is evaluated using three measurement points: 
\begin{itemize}
    \item \textbf{No Transport:} For this reference measurement, no SENSEI analysis adapter is enabled in the SENSEI runtime XML configuration. However, SENSEI is still used for the measurement.
    \item \textbf{Checkpointing:} The SENSEI endpoint is configured to write the pressure and velocity fields to the storage system as VTU files.
    \item \textbf{Catalyst:} The SENSEI endpoint receives the data from NekRS-SENSEI and renders two images using ParaView over Python.
\end{itemize}

\begin{figure}[h]
\includegraphics[width=0.9\linewidth]{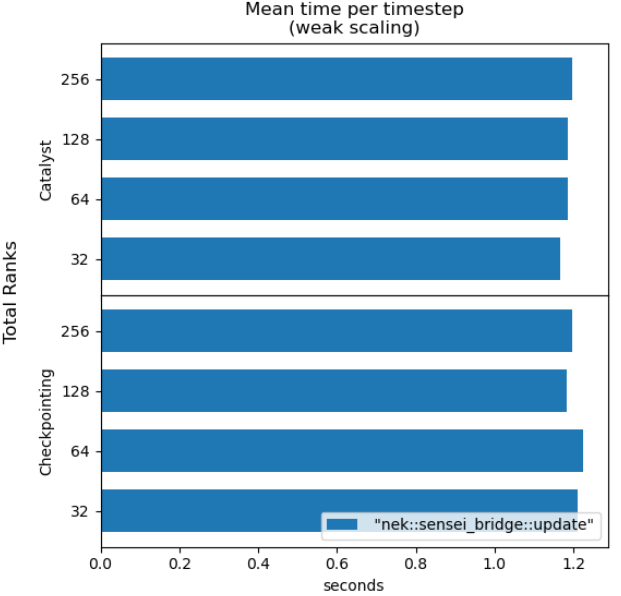}
\caption{Measurement of the mean time per timestep on the NekRS-SENSEI simulation nodes. Each rank represents one GPU.}
\label{fig:jsc_time}
\end{figure}

\begin{figure}[h]
\includegraphics[width=0.9\linewidth]{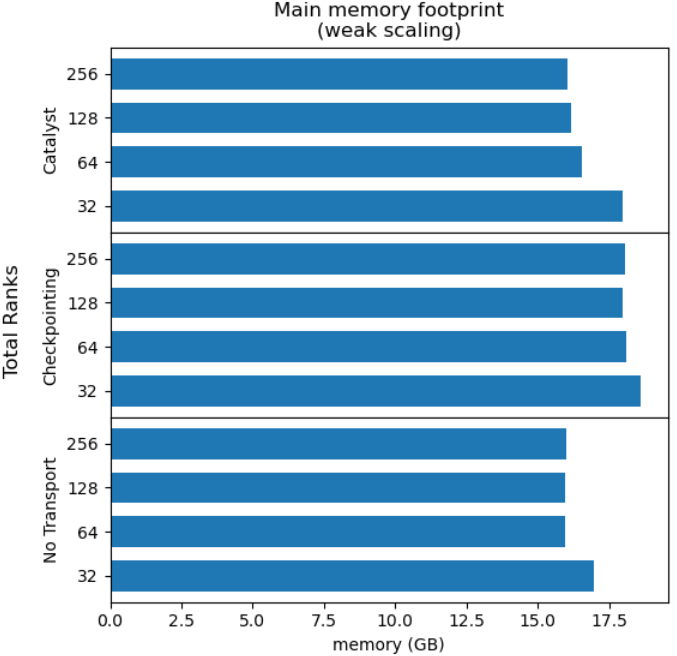}
\caption{Measurement of the main memory footprint per NekRS-SENSEI simulation node. Each rank represents one GPU.}
\label{fig:jsc_mem}
\end{figure}

As before for the \textit{in situ} case, our analysis for the in transit case focuses on the overhead in terms of time (Fig.~\ref{fig:jsc_time}) and memory (Fig.~\ref{fig:jsc_mem}) caused by the visualization for the simulation node. Furthermore, the scalability is evaluated by means of weak scaling, i.e. the theoretical load per node is kept constant as the number of nodes is increased. 


The measurements shown in Figure~\ref{fig:jsc_time} highlight two important points for the in transit workflow. First, the times for the Catalyst and Checkpointing measurement points are very similar, meaning that the in transit overhead is small. Second, the times for increasing nodes do not increase significantly. Therefore, weak scaling works well, and it can be assumed that the workflow will work for even larger setups.

This result is also supported by Figure~\ref{fig:jsc_mem}. The memory consumption for Catalyst and No Transport is very similar. The memory overhead of Checkpointing is visible, but not very large. Again, note that the memory available for the simulation nodes is independent of the number of visualizers.

\section{Discussion and Conclusion}
Across both the \textit{in situ} Pebble-bed reactor case and the \textit{in transit} Mesoscale case, we have evaluated the computational paradigms that leverage contemporary \textit{in situ} and in transit visualization methodologies. This exploration aimed to balance computational efficiency, storage necessities, and effective visualization in the dynamic landscape of high-performance computing.

A recurring theme in our assessments is the value that advanced visualization brings to computational workloads. Whether analyzing the flow dynamics within a pebble-bed reactor or exploring the turbulent flows of Rayleigh-Bénard convection, the ability to "see" the data in real-time significantly augments our analytical capacities. Both case studies underscore the importance of managing computational efficiency in conjunction with data storage and visualization requirements. The Catalyst approach in the Pebble-bed reactor case presented an impressive storage economy, demonstrating that efficient visualization doesn't need to come at the expense of increased storage demands. Similarly, the in-transit approach for the Mesoscale case showed that with the right tools and configuration, overheads can be minimized, preserving the sanctity of computational resources. One of the key markers of success for any high-performance computing methodology is scalability. Our results, especially in the context of the in-transit visualization of Rayleigh-Bénard convection, attest to the fact that these approaches are not merely academic exercises but are scalable, efficient, and ready for the rigors of future complex simulations.

In conclusion, as the gap between I/O and computational demands widens, methodologies like \textit{in situ} and in transit analysis and visualization offer promising pathways. Our findings demonstrate that with careful configuration, integration of advanced tools, and an understanding of the underlying phenomena, we can achieve computational efficiency without compromising on visualization efficacy. As computational simulations become even more intricate and demand increased resources, such approaches will be pivotal in advancing scientific understanding.

\begin{acks}
This work was supported by and used resources of the Argonne Leadership Computing Facility, which is a U.S. Department of Energy Office of Science User Facility supported under Contract DE-AC02- 06CH11357. This work was supported by Northern Illinois University. This work was supported in part by the Director, Office of Science, Office of Advanced Scientific Computing Research, of the U.S. Department of Energy under Contract DE-AC02-06CH11357, through the grant “Scalable Analysis Methods and In Situ Infrastructure for Extreme Scale Knowledge Discovery”, program manager Dr. Margaret Lenz. The authors from JSC acknowledge computing time grants for the project TurbulenceSL by the JARA-HPC Vergabegremium provided on the JARA-HPC Partition part of the supercomputer JURECA at Jülich Supercomputing Centre, Forschungszentrum Jülich, the Gauss Centre for Supercomputing e.V. (www.gauss-centre.eu) for funding this project by providing computing time on the GCS Supercomputer JUWELS at Jülich Supercomputing Centre (JSC), and funding from the European Union’s Horizon 2020 research and innovation program under the Center of Excellence in Combustion (CoEC) project, grant agreement no. 952181. Support by the Joint Laboratory for Extreme Scale Computing (JLESC, https://jlesc.github.io/) for traveling is acknowledged.
\end{acks}

\bibliographystyle{ACM-Reference-Format}

\bibliography{paper}

\end{document}